\documentclass[onecolumn,epsfig,pre]{revtex4}
\usepackage{graphicx}
\usepackage{amssymb}
\usepackage{amsmath}
\usepackage{hyperref}
\usepackage{latexsym}

\def\be{\begin{equation}}
\def\ee{\end{equation}}
\def\bea{\begin{eqnarray}}
\def\eea{\end{eqnarray}}

\draft

\begin{document}

\title{Scale invariant pattern in dynamically extending lattice}
\author{Sudipto Muhuri }


\affiliation {Department of Physics, BITS-Hyderabad Campus, Shameerpet, Hyderabad 500078, India}
\begin{abstract} 
We study the implications of coupling Langmuir kinetics (LK) process with the dynamics of an extending lattice. The model that we consider couples  dynamically extending exclusion process (DEEP) with the  process of random attachment and detachment of particles in the bulk of the lattice. We explore a dynamical regime where the boundary processes of lattice extension and particle input compete with the bulk process of particle attachment and detachment. This competition leads to scale invariant density profile of particles when expressed in terms of the relative position in the growing lattice.  It also leads to phase coexistence and  shocks in the bulk of the lattice. We use a combination of Mean Field (MF) analysis and Monte Carlo simulations to characterize the density profile in growing lattice and construct a MF phase diagram. This study will have possible implications for transport and patterning in growing fungal filament.
\end{abstract}








\maketitle
An intrinsic feature of the biological world  is the process of growth. One fascinating aspect of growth process during embryonic development is the formation of  patterns.  
Positional information associated with pattern formation is initiated by concentration gradient of morphogens \cite{ scalingnature, morpho1}.
Experimental studies such as the one done with wild-type D. {\it melanogaster} embryos have indicated that the {\it bicoid} morphogen distribution scales linearly with the size of the embryo \cite{morpho1}, which would imply that as a function of the relative position in the embryo, the morphogen distribution is invariant.   
In general such scaled patterning is seen to be robust against environmental variations and genetic background variations \cite{morpho1}. A natural question that arises in this context is - what are the mechanisms by which such scale invariant patterning is achieved ? The answer to this question is an involved one, which requires careful consideration depending on the specific biological context.  While we are not going to address that issue in this letter, it would serve as the broad motivation for focusing our attention on a simple one dimensional lattice model for growth which exhibits such scaled growth pattern. For such scaled growth pattern, the density profile remains invariant in the scaled position variable $x/L$, where $x$ is the position and $L$ the length of the lattice. This model that we would discuss, incorporates some of the ingredients involved in  quasi one-dimensional biological growth process of individual fungi filament.

Fungi are eukaryotes and can typically grow as  cylindrical tubes (hyphae), with the growth happening at the tip of fungi \cite{deacon}. 
The material required for growth is packaged as vesicles that are  supplied to the growing tip by molecular motors. The molecular motors  transport these materials using the microtubules tracks to the growing end of the hyphae. These vesicles bring fresh membrane, and regulators neccesary for growth. The growth process of fungal hyphae have a sense of persistence in one direction. In addition to linear growth of the filament, lateral branching of hypha also appears  giving rise to a network termed as {\it mycelium} \cite{deacon}.  
Fungal mycelium exhibits patterning and it has also been observed that the spatial pattern associated with morphogenesis in fungi during spore germination requires the presence of microtubules \cite{harris}.

We model the growing fungal filament  as a one dimensional lattice which can grow at the one of the ends of the lattice.
 The model that we study belongs to the class of driven lattice gas models and a generalization of the model of dynamic lattice first proposed and studied in \cite{sugden1,sugden2}.  Driven lattice gas models have been used to model a number of biological processes, such as the motion of ribosomes in m-RNA~\cite{mcdonald}, motor and vesicle transport~\cite{freypre, menon, ashwin, ignaepl,ignapre}, transport across biomembranes~\cite{choubio} and growth of fungal hyphae \cite{sugden1,sugden2}. Many of these models are based on the totally asymmetric exclusion process (TASEP), in which the particles jumps unidirectionally on the lattice  with a single rate and interacts with other particle via hard-core repulsion. For TASEP, the entire non-equilibrium phase diagram in terms of particle fluxes at the boundaries, has been worked out exactly\cite{evans}. 
TASEP has been generalized to incorporate a dynamically extending lattice, which is growing at one the ends of the lattice \cite{sugden1,sugden2}. For such dynamically extending TASEP, the growth velocity at tip is proportional to the particle density at the tip and this minimal model has been proposed to describe the growth of fungal hyphae \cite{sugden1,sugden2}. Number of other kinds of dynamic lattice gas models have  studied in varied context such as a dynamic lattice coupled to a diffusive reservoir \cite{santen1,sugdenthesis, santen, chou,kafri} and a dynamic lattice  which includes process of corrosion\cite{platini}.

Here we consider the coupling of the lattice growth process with Langmuir kinetics. For the case of growth of fungal hyphae, the fact that the transported cargo may detach (and reattach) to the microtubule filament makes it plausible scenario to explore. In particular, we study the dynamic regime where these two processes compete; the kinetic rates are such that fluxes of the particles at the boundaries are comparable to that fluxes in the bulk. For a static lattice, the competition beween particle fluxes at the boundaries with the process of particle exchange with the surrounding bath has been addressed in \cite{freylet}. We study this competition in context of dynamically extending lattice, focussing on a regime where the site attachment and detachment rates scale inversely as the system size. We use mean field (MF)  analysis  to study the system and compare these results with those obtained by perfoming Monte Carlo simulations. We find that  the competition between the growth process and kinetics of attachment and detachment not only leads to the possibility of phase coexistence and shocks in the co-moving frame of the growing lattice, it also implies scale invariant density profile in the growing lattice. This dynamic regime exhibits qualitatively new scenario when compared to the situation when the growing lattice is coupled to a bulk reservoir where the  total attachment and detachment rates scales extensively with the filament size \cite {sugdenthesis}.

\begin{figure}[h]
\centering
\includegraphics[width= 2.4in,height = 0.7in,angle=0]{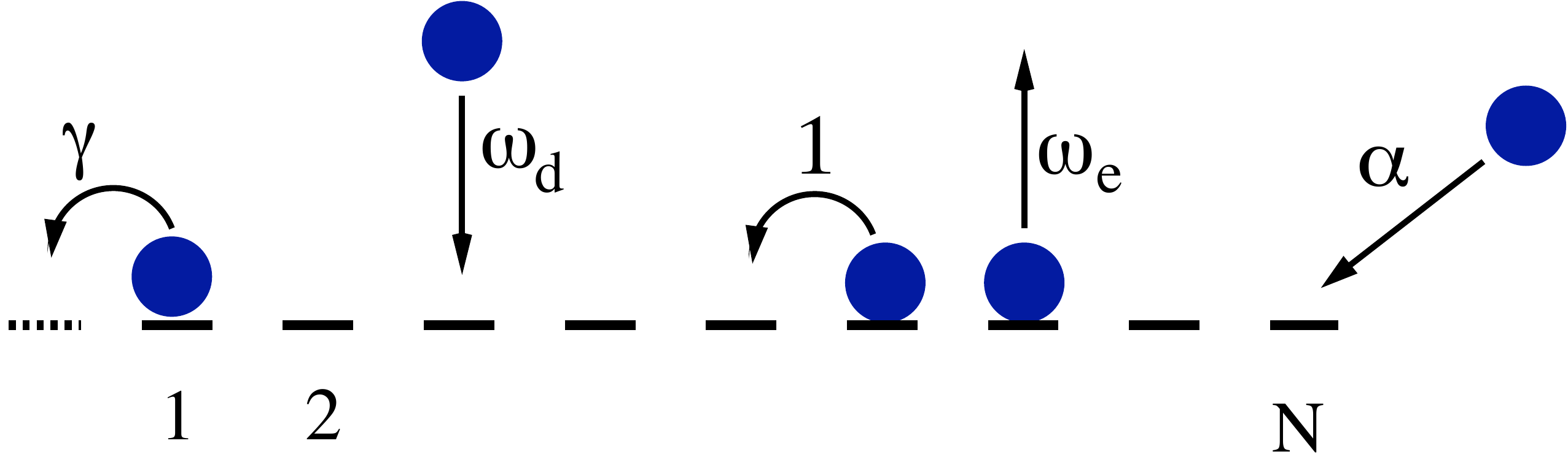}
\begin{center}
\caption{Schematic representation of the dynamical processes on the growing lattice with particle entry rate $\alpha$, growth rate $\gamma$, attachment rate $\omega_{d}$, detachment rate $\omega_{e}$ and hopping rate $1$.}
\label{fig1}
\end{center}
\end{figure}

We represent the growing filament at any instant of time to be a one-dimensional lattice of length L and lattice sites N with a lattice spacing  $\epsilon = L/N$. Particles hop unidirectionally to the left  on this lattice with the instantaneous configuration being described in terms of  the occupation number of particles at each lattice site $i$, which can assume values   $n_{i} =1$ if the site $i$ is occupied or $n_{i}= 0$ if the site is vacant. The lattice length and the number of lattice sites increase due the growth process which happens at the left end of the filament. The lattice growth happens when a particle moving unidirectionally towards the left on the lattice, reaches the filament end and creates  a new lattice site with a rate $\gamma$ so that the total lattice length increases by an amount  $\epsilon$. The non-growing right end of the  filament  is supplied with particles which enters the rightmost site with a rate $\alpha$. In the bulk of the lattice two distinct processes can occur:  particles can hop to the neighboring site on the left  with rate $1$,  if the site is vacant;  particles can detach from the filament  with rate $\omega_{e}$ or a new particle can  attach to the filament  lattice site with a rate $\omega_{d}$ provided the site is vacant.  The various dynamical processes are summarized in Fig.1. We focus on the situation, similar to \cite{freylet}, where the overall (un)binding rates for the entire lattice is held fixed, so that the site (un)binding rates vary inversely with the system size. 
Since the lattice size continously increases, there are two frames to consider; while one is the laboratory frame, the other is a frame attached to the growing tip. We choose the reference frame attached to the growing tip of the filament, with the leftmost site being labelled as $1$ and the rightmost being labelled as site $N$.  We write the evolution equation of the occupation number of particles in the aforementioned frame.  Any site $i$ measures the distance of that site from the growing tip. Each lattice site can be occupied by at most one particle.  Whenever the lattice grows due to creation of a new site at the tip, all the site labels are updated $i\rightarrow i+1$. 
The evolution equation for the occupation numbers in the bulk at site $i$ is,
\begin{eqnarray}
\frac{\partial n_{i}}{\partial t}&=& n_{i+1}(1-n_{i}) -n_{i}(1-n_{i-1}) +
\gamma n_{1}(n_{i-1} -n_{i})\nonumber\\& -&\omega_{e}n_{i} + \omega_{d}(1 - n_{i}),   ~~~~~~~~~~i\geq3
\end{eqnarray}
The terms on the right hand side of the equation have their usual interpretation of gain and loss terms due to translation of the particles, the effect of the growing tip and the Langmuir kinetics. 
The corresponding current at  site $i$ can be identified as, 
$J_{i} = ( 1 - n_{i-1})n_{i} - n_{1}n_{i-1}$.
The occupation number of the particles at sites 1 and 2  in the comoving frame of the tip are respectively,
\begin{eqnarray}
\frac{\partial n_{1}}{\partial t}&=& -\gamma n_{1} + n_{2}(1-n_{1}) 
\label{eq:mfbound1} 
\\
\frac{\partial n_{2}}{\partial t}&=& n_{3}(1-n_{2}) -n_{2}(1-n_{1}) - \gamma n_{1}n_{2} -\omega_{e}n_{2} + \omega_{d}(1 - n_{2})\nonumber
\label{eq:mfbound2} 
\end{eqnarray}
At the right end of the filament  the evolution equation reads,

\begin{equation}
\frac{\partial n_{N}}{\partial t} = \alpha (1-n_{N}) -n_{N}(1-n_{N-1}) 
\label{eq:rb1}
\end{equation}

Defining $\rho = \langle n_{i} \rangle$, we introduce mean field approximation which amounts to taking averages over ensembles and factorizing the two-point correlators, $\langle n_{i}n_{i+1}\rangle $ =  $\langle n_{i}\rangle \langle n_{i+1}\rangle $ \cite{privman,ignapre}. 
We now write the continuum evolution equation in terms of the rescaled variable x = $i\epsilon/L$, which measures the relative position of the particle in the growing lattice.
We define the reduced rates, $\Omega_{d} = N\omega_d$ and $\Omega_{e} = N\omega_e$. These rates have the interpretation of the total bulk attachment and detachment rates of particles for the entire growing lattice. In the thermodynamic limit of $N\rightarrow \infty$, we obtain the  the continuum evolution equation as, 

\begin{equation}
\frac{\partial \rho}{\partial t} =  \frac{\partial}{d x}[\rho(1-\rho)]- \gamma n_{1}\frac{\partial \rho}{dx}  -\Omega_{e}\rho +\Omega_{d}(1 -\rho) 
\label{eq:mfbulk}
\end{equation}
where we have retained terms upto leading  order in $1/N$.

In the continuum description,  the evolution equation of density at the  non-growing end at $x =1$ satisfies,

\begin{equation}
\frac{\partial \rho(1)}{\partial t} = \alpha[1-\rho(1)] -\rho(1)[1-\rho(1)]
\label{eq:rb} 
\end{equation}

The evolution equation for particle densities at sites $1$ and $2$ are respectively,
 
\begin{eqnarray}
\frac{\partial \rho_{1}}{\partial t}&=& -\gamma \rho_{1} + \rho_{2}(1-\rho_{1}) 
\label{eq:bound1} 
\\
\frac{\partial n_{2}}{\partial t}&=& \rho_{3}(1-\rho_{2}) -\rho_{2}(1-\rho_{1}) - \gamma \rho_{1}\rho_{2} 
\label{eq:bound2} 
\end{eqnarray}
where $\rho_{1} = \langle n_{1}\rangle $ and  $\rho_{2} = \langle n_{2}\rangle$.
In order to determine the steady state density at the boundary of the growing tip, we make the approximation,
\begin{equation}
 \rho_{3} = \rho_{2}
 \label{eq:approx}
 \end{equation}
 
Then using Eq.(\ref{eq:bound1}) and Eq.(\ref{eq:bound2}) we obtain, 
\begin{eqnarray}
\rho_{1} &=& \frac{1- 2\gamma}{1 -\gamma}
\label{eq:rho1a}
\\
\rho_{2} &=& 1 - 2\gamma
\label{eq:rho2}
\end{eqnarray}

The approximation of Eq.(\ref{eq:approx}) works reasonably well,  when the bulk  phase adjoining the tip of the growing tip  is a high density phase. So for example if the bulk phase is a high density phase or even a mixed phase of high and low density, then this approximation can predict the steady state density profile at the growing tip boundary reasonably well, as seen by comparison with profiles obtained by Monte Carlo simulations.  We shall elaborate more on the domain of validity of this approximation, when we  discuss the possible phases and construct the phase diagram later.

Having obtained the the expression for $\rho_1$, the continuum steady state equation assumes the form, 
 
\begin{equation}
\frac{dJ}{dx}   - (\Omega_{d} + \Omega_{e})\rho + \Omega_{d} = 0 
\label{eq:ss1}
\end{equation}
where,
\begin{equation}
J = \rho(1-\rho) - \frac{\gamma(1-2\gamma)}{1 - \gamma} \rho
\label{eq:J}
\end{equation}
J has the interpretation of particle current in the bulk of the lattice in the frame of the moving tip.

Integrating Eq.(\ref{eq:ss1}), we obtain,
\begin{equation}
2(\rho_{b} - \rho) + \left[ 1 - \frac{\gamma(1 - 2\gamma)}{1 - \gamma} - \frac{2\Omega_{d}}{\Omega}\right]\log\left[\frac{ \Omega\rho  -\Omega_{d}}{\Omega\rho_{b}  - \Omega_{d}} \right]= \Omega( x - x_{b})
\label{eq:ss}
\end{equation}

where $\rho_{b}$ is the boundary density and $x_{b}$ is the position of the boundary;  $x_{b} =0$ for the left boundary and $x_{b}=1$ for the boundary at the right end of the filament. Here $\Omega = \Omega_{d} + \Omega_{e}$. 

\begin{figure}[h]
\centering
\includegraphics[width= 3in,height = 2.2in,angle=0]{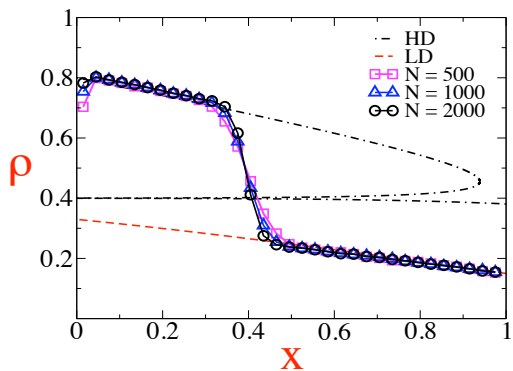}
\begin{center}
\caption{Steady state density profile of particles in scaled variable $x$. Here, $\alpha = 0.15, \gamma = 0.1, \Omega_{d} = 0.2$ and $\Omega_{e} = 0.3$. The square, triangle and circles  with solid lines represent the density profile obtained by Monte Carlo numerical simulations for different sizes of the growing lattice. The dashed lines are the MF densities solutions corresponding to low density(LD) and high density(HD) phases obtained from Eq. (\ref{eq:ss}). }
\label{fig2}
\end{center}
\end{figure}

When the density profile satisfies the boundary condition  at $x =0$, then  the boundary density is obtained by equating to the occupation number density at site $2$ to the boundary density, so that $\rho(0) = \rho_{2} = 1 - 2\gamma$. This density profile corresponds to a High density (HD) phase. For  the density profile  which satisfies the boundary condition at the other end of the filament  located at  $x =1 $, we obtain  $\rho(1) = \alpha$. This density profile corresponds to a Low density (LD) phase.
In general, the boundary conditions imposed at the two filament ends cannot be satisfied simultaneously. As a result,  at least two domains usually  develop along the filament, giving rise to the possibility of shock profiles. The location of the shock is determined by equating the current  $J$ for the LD and HD solution at the position of the shock.  Fig. \ref{fig2} shows shock density profile obtained by Monte Carlo simulations for different system sizes. These profiles are compared with the MF prediction of Eq.(\ref{eq:ss}) corresponding to the LD and HD  mean field solution of density.  As seen in Fig.\ref{fig2},   the entire density profile including the location of shocks overlaps for the different system size in the scaled variable $x$. This would imply that the  growth process is such that the density profile remains invariant for different system sizes in the scaled variable $x$. We also observe that the MF predictions matches reasonably well with the Monte Carlo simulation results.
In order to analyse the different dynamical growing phases that are possible for this system, we focus on a special regime where the kinetic rates satisfy the constraint, 
\begin{equation}
\frac{1}{2}(1 -  \gamma \rho_{1})=\frac{\Omega_{d}}{\Omega_{e} + \Omega_{d}} =   K_{d}
\label{eq:lincond}
\end{equation}
 For this choice of constraint, the steady state equation for the bulk reduces to,
\begin{equation}
\left[\rho - \frac{\Omega_{d} }{\Omega_{d} + \Omega_{e}}\right]\left[\frac{d\rho}{dx} + B\right] = 0 
\label{eq:mflinear}
\end{equation}
where $B= (\Omega_{d} + \Omega_{e})/2$.\\
Thus for this particular choice of constraint, the steady state MF density profiles either vary  linearly  with $x$ or are homogenous with respect to $x$. As a result,  there are three different profiles which may coexist along the filament. The different solutions correspond to a linearly decreasing high density (HD) solution which satisfies the boundary density  at the growing tip $( x = 0)$, a linearly decreasing low density (LD) solution which satisfies the boundary condition at $x=1$  and a homogenous density solution (U)  which is similar to a density profile determined by Langmuir kinetics alone. The  non-equilibrium phase diagram can be constructed once we determine the limits of coexistence of the phases.  Phase coexistence in bulk can occur when the densities in the linear phases (LD or HD)  matches with the density of the homogenous solution (U)  at $x = x_{o}$, in the domain range  $0 < x < 1$,  where  $x_{o}$ is position of the domain wall separating the two phases. For  LD phase to coexist with the HD phase, current of the LD phase must equal the particle current in the HD phase at  domain wall position $x=x_{o}$ in the domain $0 < x < 1$. Phases on the either side of the domain wall would  have a density profile corresponding to the particular phase on the growing lattice. The phase boundaries are determined by the condition of expulsion of one of the phases from the bulk, i.e; by setting the location of the domain wall position to the filament boundaries. While the density changes continuously on moving across the phase boundary separating the U region with the HD(LD) region, the density change is dicontinuous while moving from a LD region to a HD region. 

In order to analytically determine the MF density profile and the phase boundaries, separating the different phase regions, we again use the approximation  introduced in Eq.(\ref{eq:approx}),  so that  the expressions for $\rho_{1}$ and $\rho_{2}$ are determined by Eq.(\ref {eq:rho1a}) and Eq.(\ref{eq:rho2}), as before. This approximation scheme implicitly assumes that $\rho_{1}$ is determined by the boundary conditions at the growing end of the filament alone. This is valid when the bulk density adjoining $x=0$ is determined by the growth rate at growing top alone,  and  the rate of tip growth is the rate limiting process which controls the bulk density adjoining the region of the growing tip.  However when entire bulk phase is a LD phase and  the HD phase is totally expelled from the growing end of the filament, then this approximation will not hold good.  As long as the high density phase is present in the bulk, this approximation is reasonable, as confirmed by Monte Carlo simulations and illustrated in Fig.2. The condition for linear solution is then,
\begin{equation}
K_{d} = \frac{1}{2}\left[1 -  \frac{\gamma(1 - 2\gamma)}{1 - \gamma}\right]
\label{eq:condlin}
\end{equation}
and the corresponding expressions for MF density profile for the LD, U and HD phases are respectively,

\begin{equation}
\rho(x)=\left\{\begin{array}[c]{ll}  \rho_{hd}(x) = 1 - 2\gamma + Bx  & \\
 & \\
\rho_{u}(x) = \frac{1}{2} -\frac{\gamma(1 -2\gamma)}{2(1 -\gamma)} & \\
 & \\
\rho_{ld}(x) = \alpha + B(1-x) &
\end{array}
\right.
\end{equation}

By varying the parameters $\gamma$ and $\alpha$,  for a fixed value of $\Omega_{d}$, one can see the existence of both single phase regions, regions of phase coexistence and presence of  shocks in the bulk. 

\noindent
{\it Phase coexistence line between the HD and LD-HD:} This is determined by using the expression for current  $J$ in Eq.(\ref{eq:J}) and matching the current  for the HD phase, with the current in LD phase at $x=1$, so that $J_{ld}(1) = J_{hd}(1)$.  We thus obtain the equation of phase boundary as,
\begin{equation}
\alpha^{2} - \frac{1 - 2\gamma +2\gamma^{2}}{1- \gamma}\alpha + J_{HD}(1) = 0
\end{equation}
 where, $J_{hd}(1) = (1-2\gamma - B)(2\gamma + B) -\frac{\gamma(1-2\gamma)(1 -2\gamma - B)}{1-\gamma}$.
 
\noindent
{\it Phase coexistence line between the LD and LD-HD:} On matching the current for the HD phase, with the current in LD phase at $x=0$, so that $J_{ld}(0) = J_{hd}(0)$, we obtain the equation of phase boundary as,
\begin{equation}
\alpha^{2} - \left[ 1 - 2B - \frac{\gamma(1 -2\gamma)}{1- \gamma}\right] \alpha + J_{hd}(0) + B^{2} - B\frac{1-2\gamma + 2\gamma^{2}}{1-\gamma} = 0
\end{equation}
 where, $J_{hd}(0) = \frac{\gamma(1-2\gamma )}{1 - \gamma} $.
 Here we have determined the location of phase boundaries in the phase diagram using Eq.(\ref{eq:condlin}). This maybe understood in the following manner:  Consider a region in the vicinity of the left boundary with $x=\delta$. Let the region on the right of $x$ be a LD phase while the region on the left (from $x= 0$ to $x=\delta$) be in the HD phase. For this situation, condition for the phase boundary separating the LD phase with LD-HD phase coexistence region is obtained by letting the domain wall position $x_{o} \rightarrow 0$.  For a finite value of $\delta$, $\rho_{1} = \frac{1- 2\gamma}{1 - \gamma}$ and the condition of linearity for the coexisting LD and HD profile is set by it. Consequently if one approaches $\delta \rightarrow 0$ from the right, the boundary density for site 1 and 2 can be obtained using Eq.(\ref{eq:rho1a}) and Eq.(\ref{eq:rho2})  and  the condition for linear profile is set by Eq.(\ref{eq:condlin}). Since the phase boundary can be  obtained by the limiting procedure of setting the domain wall position $x_{o}$ to  $x= 0$, approaching from the right, thus the phase boundary itself may be  determined using  the same condition of linearity as Eq.(\ref{eq:condlin}) and value of $\rho_{1}$ is determined by eq.(\ref{eq:rho1a}) although the LD phase on the otherside of the phase boundary in the phase diagram would not satisfy the same condition for linear solution as before. 

\noindent
{\it Phase coexistence line between the LD-HD and LD-U-HD:} 
If we require that the density for LD phase is same as that of  U phase at the domain wall location, then the expression for position of the domain wall is $ x_{ld/u} = \frac{\alpha + B - \rho_{u}} {B}$. Similarly expression for the domain wall position separating HD with U region is $x_{hd/u} = \frac {1-2\gamma - \rho_{u}}{B}$. The condition for phase coexistence boundary is that $x_{ld/u} = x_{hd/u}$. Thus we obtain the equation for phase boundary as,
\begin{equation}
\alpha = 1 - 2\gamma -\frac{\Omega_d}{2}\left[\frac{2 -2\gamma - \gamma^{2}}{1 -2\gamma + \gamma^{2}}\right] 
\end{equation}

\noindent
 {\it Phase coexistence line between the HD and  HD-U:} By matching the density  for the HD phase with the density in U phase at $x=1$,  we obtain the equation in terms of $\gamma$,
\begin{equation}
2\gamma^{2} + (4 - 2B)\gamma + 1 - 2B  = 0
\label{eq:pb7}
\end{equation}
Since $ B = \frac{\Omega_{d} + \Omega_{e}}{2} = \frac{\Omega_{d}(1-\gamma)}{1-2\gamma + 2\gamma^{2}}$, we numerically solve Eq.(\ref{eq:pb7}) to determine the value of $\gamma$ corresponding to the phase boundary.

\noindent
{\it Phase coexistence line between the U and HD-U:} This is determined by matching the density  for the HD phase, with the density in U phase at $x=0$. Thus the  equation of phase boundary is,
\begin{equation}
\gamma = 1 - \frac{1}{\sqrt 2} 
\label{eq:mmh}
\end{equation}

Thus we see that above this  critical value of $\gamma$ determined by Eq.(\ref{eq:mmh}), the high density phase is expelled from the bulk of the lattice. In such situation the only possibilities are  that of  homogenous $U$ phase, a pure $LD$ phase and  $LD-U$ phase coexistence.    

For a pure LD phase, the rate limiting process is the input rate $\alpha$, so that the linear profile, $\rho(x) = \alpha - B(1-x)$ would occupy the entire bulk of the lattice, right upto the growing end of the filament. In such circumstances, in order to determine the boundary value of density at $x=0$, we use  the approximation,

\begin{equation}
 \rho_{2} = \alpha + \frac{1}{2}(\Omega_{d} + \Omega_{e}) = \alpha_{o}
 \label{eq:ap2}
 \end{equation}
  This approximation also would imply that  $\rho_{1} =\frac{\alpha_{o}}{\gamma +\alpha_{o}}$. Thus the corresponding  linear solution for the low density profile would then have to satisfy the condition, 

\begin{equation}
K_{d} = \frac{1}{2}(1 - \frac{\gamma \alpha_{o}}{\gamma +\alpha_{o}})
\end{equation}
This condition for linear solution is  different from the condition of linearity expressed in Eq.(\ref{eq:condlin}), where the HD phase was present in the bulk.
\begin{figure}[h]
\centering
\includegraphics[width= 2.8in,height = 2.8in,angle=0]{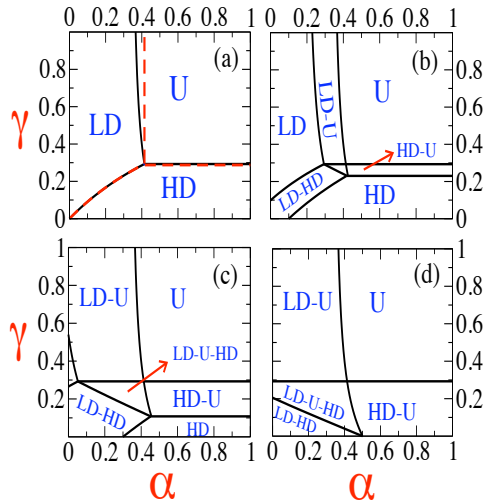}
\begin{center}
\caption{Phase plane cut along  the $\alpha- \gamma$ plane on varying the overall attachment rate of particles. (a)$\Omega_{d} = 0$: This corresponds to the situation where Langmuir kinetics is switched off. The dashed curves are the MF results derived in Ref.\cite {sugden2}. (b) $\Omega_{d} = 0.1$: Regions of phase coexistence are seen. (c) $\Omega_{d} = 0.3$ : The LD-U region widens towards the left while  HD-U region expands downwards and LD phase region is almost expelled. (d) $\Omega_{d} =0.5$ : HD phase is expelled from the phase diagram. The location of (U)-(LD-U) and (U)-(HD-U) phase coexistence line remains unchanged on variation of $\Omega_{d}$ in all cases.}.    
\label{fig3}
\end{center}
\end{figure}    

\noindent
{\it Phase coexistence line between the LD and LD-U:} This is determined by matching the density  for the LD phase, with the density in U phase at $x=0$. The value of density for the LD mean field solution at $x=0$ is, $\rho_{ld}(0) = \alpha + \frac{1}{2}(\Omega_{d} + \Omega_{e})$. The homogenous solution which determines the U phase density is given by $\rho_{u}(0) = \frac{1}{2}(1 -\gamma\rho_{1})$. We use the approximation of Eq.(\ref{eq:ap2}) and the steady state condition corresponding to Eq.(\ref{eq:bound1}), to  obtain the corresponding expression for the $\rho_{u}(0)$. By equating the expressions for the densities in the two phases, we obtain we density at the left boundary as $\rho  = \frac{1}{4} [1 - 3\gamma + \sqrt{\gamma^{2} + 2\gamma + 1}]$, and the equation for the phase coexistence boundary is,
\begin{equation}
\alpha = \rho  -\frac{\Omega_{d}(\gamma + \rho)}{\gamma + \rho - \gamma\rho}
\end{equation}
The corresponding condition for linear solution is,
$\alpha= K_{d} - \frac{1}{2\Omega_{D}K_{d}}$.

\noindent
 {\it Phase coexistence line between the U and LD-U:} The coexistence boundary is determined by matching the density  for the LD phase with the density in U phase at $x=1$. For this case $\rho_{ld}(1) = \alpha$ and $\rho_{u}(1) = \frac{1}{2}( 1 -\gamma\rho_{1}) = \frac{1}{2}(1 - \frac{\gamma\alpha}{\gamma + \alpha})$, where we have again made use of the approximation, 
 \begin{equation}
 \rho_{2} = \alpha
 \label{eq:ap3}
 \end{equation}
 
 Thus the  equation of phase boundary is,
\begin{equation}
\gamma = \frac{2\alpha^{2} - \alpha}{1 - 3\alpha} 
\end{equation}
The corresponding condition for linear solution is,
$\alpha = K_{d}.$
The MF phase diagram in Fig. \ref{fig3} displays the phase space cut along  $\alpha-\gamma$ plane  for a special choice of parameter values which are consistent with the constraint requirement specified by Eq.(\ref {eq:lincond}). Fig.\ref{fig3} illustrates how with the increase of $\Omega_{d}$, regions of phase coexistence in the bulk is expanded. 
When the Langmuir kinetic process is switched off, then the dynamics of the model  reduces to the case of a dynamic TASEP studied in Ref.\cite{sugden2}.  Fig.~\ref{fig3}(a) compares the phase boundaries obtained for the case when $\Omega_{d} , \Omega_{e} =0$ with the results obtained in Ref.\cite{sugden2} for dynamic TASEP . While the HD-U and LD-HD phase boundary obtained in this case  is exactly the same as that obtained in  Ref.\cite{sugden2},  the location of  LD-U phase coexistence lines differ due to the approximations  made through Eq.(\ref{eq:ap3}).
Although the aspects of phase coexistence and presence of density shocks in the bulk, have many similarities with the phase diagram for TASEP-LK process for static lattice \cite{freypre}, the topology of the phase diagram in this case differs significantly. Comparison of the results obtained by this approximate MF analysis with Monte Carlo simulations reveal that while  their is  broad qualitative agreement for the density profiles and the rough location of the phase boundaries, the MF density profile  does not exactly match with the actual density profile obtained by Monte Carlo simulation. The MF analysis is also not  able to accurately determine the exact location of the phase boundaries. However the feature of scale invariant density profile predicted by the MF analysis is robust as seen by comparison with Monte Carlo simulations done for different system sizes.  

 In summary, we have presented a model of dynamically extended lattice which incorporates the process of Langmuir kinetics and exhibits feature of scale invariant density profile in the bulk. It also rationalizes in simple terms the interplay of the various competing processes which affect growth. In particular it highlights the dynamic regime for which the bulk process of Langmuir kinetics competes with the boundary processes of particle input and growth at the tip. The resultant phase diagram shows rich behavior, which includes  different inhomogeneous phases, phase coexistence in bulk and the possibility of density shocks.  This model will have significant ramification in modeling transport and growth in fungal {\it hyphae}. The emergence of  scale invariant patterned growth which can be altered by the interplay of internal regulatory mechanisms  with environment effects, makes it a useful minimal model which may be suitably  modified for modeling branching and patterning  in fungal {\it mycelium}. The framework that we discuss provides a natural means to link internal regulatory mechanism like vesicle supply rate  with the branching or tip growth rates which can be dependent on external cues \cite{deacon}. The possible extensions of this model can involve extending the model to include bidirectional transport along the filament and a more realistic biological scenario would also involve generalizing it to account for transport on multiple parallel tracks, as is the case for transport in fungal hyphae \cite{deacon}. It would then be interesting to study whether transport on multiple tracks can lead to phase segregation of material within the tracks as has been seen in \cite{ignajstat}.

\section{Acknowledgments}
I would like to thank G. Tripathy  for useful discussions and suggestions.

\end{document}